\pdfoutput=1
\documentclass{aa}

\usepackage{txfonts}
\usepackage{natbib}
\usepackage[colorlinks=true,linkcolor=blue,citecolor=blue]{hyperref}

\begin{document}

\defcitealias{frogner_accelerated_2020}{Paper I}

\title{Implementing accelerated particle beams in a 3D simulation of the quiet Sun}


\author{
    L. Frogner
    \and
    B. V. Gudiksen
}

\institute{
    Institute of Theoretical Astrophysics,
    University of Oslo,
    P.O.Box 1029 Blindern,
    N-0315 Oslo,
    Norway
    \and
    Rosseland Centre for Solar physics (RoCS),
    University of Oslo,
    P.O.Box 1029 Blindern,
    N-0315 Oslo,
    Norway
}

\date{}

\abstract
{
    The magnetic field in the solar atmosphere continually reconnects and accelerates charged particles to high energies. Simulations of the atmosphere in three dimensions that include the effects of accelerated particles can aid our understanding of the interplay between energetic particle beams and the environment where they emerge and propagate. We presented the first attempt at such a simulation in a previous paper, emphasising the physical model of particle beams. However, the numerical implementation of this model is not straightforward due to the diverse conditions in the atmosphere and the way we must distribute computation between multiple CPU cores.
}
{
    Here, we describe and verify our numerical implementation of energy transport by electron beams in a 3D magnetohydrodynamics code parallelised by domain decomposition.
}
{
    We trace beam trajectories using a Runge--Kutta scheme with adaptive step length control and integrate deposited beam energy along the trajectories with a hybrid analytical and numerical approach. To parallelise this, we coordinate beam transport across subdomains owned by separate processes using a buffering system designed to optimise data flow.
}
{
    Using an ad hoc magnetic field with analytical field lines as a test scenario, we show that our parallel implementation of adaptive tracing efficiently follows a challenging trajectory with high precision. By timing executions of electron beam transport with different numbers of processes, we found that the processes communicate with minimal overhead but that the parallel scalability is still sublinear due to workload imbalance caused by the uneven spatial distribution of beams.
}
{}


\maketitle


\section{Introduction}
\label{sec:introduction}

The interaction between plasma and magnetic field in the solar atmosphere is an efficient driver of particle acceleration. As magnetic energy, built up by convective motions, gets released through the process of magnetic reconnection, it generates a dynamic electromagnetic environment where charged particles can be accelerated. The surrounding large-scale magnetic field restricts the motion of the accelerated particles, leading to field-aligned particle beams that transport released magnetic energy away from the reconnection site and eventually deposit it into the ambient plasma. Observations of the radiation the beams directly and indirectly produce have established their crucial role in energy transport during solar flares \citep[see e.g.][and references therein]{holman_implications_2011, kontar_deducing_2011, benz_flare_2017}. Detailed numerical models of individual particle beams in isolated one-dimensional atmospheres have provided a great deal of further insight \citep[e.g.][]{somov_hydrodynamic_1981, hawley_solar_1994, liu_combined_2009}.

Less clear than the role of particle beams in isolated flaring events is how particle acceleration and transport affect, and are affected by, the overall behaviour of a complex 3D atmosphere. In \citet{frogner_accelerated_2020}, hereafter Paper I, we presented a model for acceleration and transport of energetic electrons in the context of a 3D magnetohydrodynamics (MHD) simulation. The model, which is a part of the Bifrost simulation code \citep{gudiksen_stellar_2011}, employs a simple parametric treatment of acceleration to determine the properties of the beams accelerated in reconnection regions. It then traces the trajectory of each beam away from the reconnection region where it originates and computes the energy it deposits into the ambient plasma along the way. We add this as a term in the MHD energy equation in Bifrost to enable the beams to influence the further evolution of the atmosphere.

There are several numerical challenges, not discussed in \citetalias{frogner_accelerated_2020}, that we had to solve to obtain an accurate and efficient implementation of the model. Because the number of beams generated at every time step can be of the order of several million, simulating a single beam must be very fast. At the same time, computing the beam's trajectory and energy deposition has to be accurate for the diverse range of conditions it may encounter. Finally, the implementation must fulfil these criteria in an atmosphere decomposed into subdomains distributed between numerous processes that execute the simulation in parallel.

In this paper, we discuss our solutions to these problems. We employ an embedded Runge--Kutta method with adaptive control of step length to compute beam trajectories quickly and accurately. To correctly integrate deposited energy along the trajectory, we derive an approximate energy transport equation suitable for integration analytically or with a Gaussian quadrature rule. Lastly, we utilise a buffering system that enables communication of partially transported beams between multiple processes with minimal overhead.

\section{Methods}
\label{sec:methods}

\subsection{Tracing magnetic field lines}
\label{sec:tracing}

The primary numerical challenge in implementing the electron beam transport model is tracing magnetic field lines. When a beam of accelerated electrons escapes the acceleration region, the Lorentz force gives the electrons a gyrating motion whose centre travels in the direction of the magnetic field. Because we assume that the time the electrons take to slow down to thermal speeds or leave the simulation domain is much shorter than the time scale on which the magnetic field evolves \citepalias{frogner_accelerated_2020}, we can treat an electron beam as traversing its trajectory instantly. Consequently, its trajectory aligns with an instantaneous magnetic field line. By tracing the field line, we thus determine the path along which the beam will deposit its energy.

To determine the path $\mathbf{x}(s)$ of a magnetic field line as a function of distance $s$ along it, we need to solve the following equation:
\begin{equation}
    \label{eq:field_line_definition}
    \frac{\mathrm{d}\mathbf{x}}{\mathrm{d}s} = \frac{\mathbf{B}(\mathbf{x})}{\left\lVert\mathbf{B}(\mathbf{x})\right\rVert}.
\end{equation}
The equation states that tangent of the field line, $\mathrm{d}\mathbf{x}/\mathrm{d}s$, everywhere points in the direction of the local magnetic field $\mathbf{B}(\mathbf{x})$. The general way of solving Eq. \eqref{eq:field_line_definition} numerically is to perform a sequence of steps, where at step $n$ we determine a direction $\mathbf{d}_n$ that will take us from the current position $\mathbf{x}_n$ to a new position $\mathbf{x}_{n+1} = \mathbf{x}_n + \Delta s_n\mathbf{d}_n$ with a step length of $\Delta s_n$.

A key concern for achieving this as efficiently as possible is how to determine step lengths appropriately for the magnetic field in the solar atmosphere. The magnetic field exhibits significant variation in smoothness throughout the solar atmosphere. In the tenuous corona, where magnetic forces dictate the motion of the plasma, field lines are primarily smooth. In the lower atmosphere, where the plasma motions can advect the magnetic field, field lines may be highly irregular. To avoid unnecessarily small steps in the smooth regions while still retaining enough accuracy in the irregular regions, it is best to use a solution method that can adapt the step length to the local conditions.

We applied a special class of solver known as embedded Runge--Kutta methods to enable efficient step length adaptation. The principle behind all Runge--Kutta methods is to iteratively improve a guess of the direction to step in before taking the step \citep{butcher_history_1996}. By taking a step using the current best approximation, sampling the direction at that position, and averaging it with the current approximation, we obtain a new approximation that is more accurate. A specific Runge--Kutta method is defined by the number of times it repeats this and the weights it uses for averaging directions. We can write the relation between the correct direction $\tilde{\mathbf{d}}_n$ (which would land us exactly on the field line) and its approximation $\mathbf{d}_n$ for a particular Runge--Kutta method as
\begin{equation}
    \label{eq:best_direction_error_order}
    \tilde{\mathbf{d}}_n = \mathbf{d}_n + \mathcal{O}\left({\Delta s_n}^p\right).
\end{equation}
Here, $p$ is the order of the method. Higher-order methods produce a smaller error term $\mathcal{O}({\Delta s_n}^p)$ for a given step length. For a given order and step length, the error indicates how irregular the local magnetic field is. Thus, provided we can estimate the value of the stepping error, we can use it to inform the choice of step length.

The embedded subclass of Runge--Kutta methods have the property that their stepping error can be estimated very efficiently \citep{fehlberg_low-order_1969, dormand_family_1980,shampine_practical_1986}. With these methods, we can average the set of sampled directions in such a way as to produce an approximation $\mathbf{d}_n^*$ to the correct direction that is of a lower order, $p-1$, compared to the best approximation $\mathbf{d}_n$, which is of order $p$. In analogy to Eq. \eqref{eq:best_direction_error_order}, we have
\begin{equation}
    \label{eq:lower_direction_error_order}
    \mathbf{d}_n^* = \mathbf{d}_n + \mathcal{O}\left({\Delta s_n}^{p-1}\right).
\end{equation}
The difference between the best and the lower-order direction approximation yields an estimate of the stepping error. Because no additional directions need to be sampled, estimating the stepping error this way is computationally cheap.

Once we have the stepping error, we can use it to decide if we should retry the current step with a smaller step length or adjust the length for the next step. To determine how the step length should be adjusted, we must compare the estimated stepping error to a target error. We can compute the estimated displacement between the correct point to step to and the point that we actually stepped to as
\begin{equation}
    \label{eq:stepping_error_displacement}
    \boldsymbol\delta_n = \Delta s_n\left(\mathbf{d}_n - \mathbf{d}_n^*\right).
\end{equation}
We then define a tolerance for displacement,
\begin{equation}
    \epsilon = \epsilon_\mathrm{abs} + \epsilon_\mathrm{rel} L,
\end{equation}
where $\epsilon_\mathrm{abs}$ and $\epsilon_\mathrm{rel}$ are parameters for absolute and relative tolerance, and $L$ is a length scale that we set to the height of the simulation domain. Optimally, the step length should be small enough to be within the tolerance but not much smaller, as that would be computationally wasteful. To judge how close the step length is to optimal, we define the quantity
\begin{equation}
    E_n = \frac{\left\lVert\boldsymbol\delta_n\right\rVert}{\epsilon}.
\end{equation}
If $E_n > 1$, the error is too large, and we should retry the step with a smaller step length. If $E_n < 1$, the error is smaller than necessary, and we should increase the step length for the next step. Inserting Eqs. \eqref{eq:best_direction_error_order} and \eqref{eq:lower_direction_error_order} into Eq. \eqref{eq:stepping_error_displacement}, we can determine that $\left\lVert\boldsymbol\delta_n\right\rVert$, and thus $E_n$, scales as ${\Delta s_n}^p$ (the error term in Eq. \eqref{eq:lower_direction_error_order} dominates the error term in Eq. \eqref{eq:best_direction_error_order}). From one step to the next, the error then scales as
\begin{equation}
    \frac{E_{n+1}}{E_n} = \left(\frac{\Delta s_{n+1}}{\Delta s_n}\right)^p.
\end{equation}
Since the target error for the next step is $E_{n+1} = 1$, the optimal next step length becomes
\begin{equation}
    \Delta s_{n+1} = E_n^{-1/p}\Delta s_n.
\end{equation}
In practice, this adjustment tends to be too aggressive, leading to oscillations of the error around $E_n = 1$ and thus frequent rejection of steps. One mitigation is introducing a safety factor $\sigma \approx 0.9$ that slightly reduces the adjusted step length to avoid overshooting. Another, proposed by \citet{gustafsson_api_1988}, is to apply principles from control theory to smooth out the sequence of step lengths. The resulting step length adjustment becomes
\begin{equation}
    \Delta s_{n+1} = \sigma {E_n}^{-\alpha}{E_{n-1}}^{\beta}\Delta s_n,
\end{equation}
where $\alpha$ and $\beta$ are positive parameters that, in general, should be tuned to the integration method. We obtained stable results with the values $\alpha = 0.3/p$ and $\beta = 0.4/p$ suggested as a starting point by \citet{gustafsson_control-theoretic_1994}.

When we simulate the transport of an electron beam, we must deposit the incremental energy loss of the beam into the ambient plasma in every adjacent grid cell along the trajectory. If we skip multiple grid cells, unphysical gradients in the deposited heat may occur. The simplest solution is to deposit lost beam energy at regular increments $\Delta s_\mathrm{dep}$ smaller than the extent of a grid cell. A consistently short distance between deposition points has the additional advantage of allowing us to simplify the calculation of the energy to deposit, as we discuss in Sect. \ref{sec:integrating_deposited_energy}. To control deposition point spacing independently from field line tracing, we need the ability to interpolate between the positions $\mathbf{x}_n$ and $\mathbf{x}_{n+1}$ to obtain intermediate deposition points. We achieved this by applying formulas derived for the embedded Runge--Kutta methods that can provide accurate interpolated deposition points by utilising the directions already sampled for the step from $\mathbf{x}_n$ to $\mathbf{x}_{n+1}$ \citep{horn_fourth-_1983,shampine_practical_1986,dormand_runge-kutta_1986}.

In addition to interpolating positions along field lines, it is necessary to interpolate various quantities whose values are known only at the centres or faces of grid cells. We must be able to evaluate the magnetic field at any position with better than grid cell scale precision to compute the stepping direction $\mathbf{d}_n$. In addition, we require the values of quantities such as temperature and electron density at the deposition points for computing the beam energy loss. To interpolate a quantity at a point $\mathbf{x}$, we evaluated a 3D polynomial of order $N$ fitted to the $(N+1)^3$ quantity values defined at discrete coordinates surrounding $\mathbf{x}$. When any of the surrounding grid cells would be outside the simulation domain, as happens when $\mathbf{x}$ is close to a non-periodic boundary, we used the grid cells closest to but inside the boundary for interpolation. We applied Neville's algorithm successively along each of the three dimensions to evaluate the interpolating polynomial with $\mathcal{O}(N^4)$ operations \citep{press_nevilles_2007}.

\subsection{Integrating deposited energy}
\label{sec:integrating_deposited_energy}

As we trace the trajectory of each electron beam, we compute the energy that the beam deposits into the ambient plasma due to Coulomb collisions at regular increments $\Delta s_\mathrm{dep}$. Under the simplifying assumptions discussed in \citetalias{frogner_accelerated_2020}, the energy deposition rate per distance at a distance $s$ along the trajectory is given by
\begin{equation}
    \label{eq:power_per_dist}
    \frac{\mathrm{d}\mathcal{E}}{\mathrm{d}s}(s) = C n_\mathrm{H}(s)\gamma(s)B\left(\mathrm{min}\left(\tau(s), 1\right); \frac{\delta}{2}, \frac{1}{3}\right){\tau^*(s)}^{-\delta/2}
\end{equation}
(from Eqs. (20), (23)\footnote{We note that we use the min function in Eq. \eqref{eq:power_per_dist} instead of the max function erroneously presented in Eq. (23) in \citetalias{frogner_accelerated_2020}.} and (27) in \citetalias{frogner_accelerated_2020}, adapted from \citet{hawley_solar_1994} and \citet{emslie_collisional_1978}). Here, $C$ is a factor depending on the initial properties of the beam;
\begin{equation}
    C = \frac{\pi e^4 (\delta - 2) P_\mathrm{beam}}{|\mu_0| {E_\mathrm{c}}^2},
\end{equation}
where $e$ is the elementary charge and $\delta$, $P_\mathrm{beam}$, $\mu_0$ and $E_\mathrm{c}$ are the power-law index, total power, pitch angle cosine, and lower cut-off energy of the initial electron distribution, respectively. Equation \eqref{eq:power_per_dist} also includes the local hydrogen number density $n_\mathrm{H}(s)$ and hybrid Coulomb logarithm $\gamma(s)$ (defined by Eq. (21) in \citetalias{frogner_accelerated_2020}), as well as the incomplete beta function $B$ (Eq. (22) in \citetalias{frogner_accelerated_2020}). The quantity $\tau(s)$ is the ratio of the hydrogen column depth $N(s)$ to the stopping column depth $N_\mathrm{c}(s)$ of electrons with energy $E_\mathrm{c}$:
\begin{equation}
    \label{eq:tau}
    \tau(s) = \frac{N(s)}{N_\mathrm{c}(s)} = \frac{1}{N_\mathrm{c}(s)}\int_0^s n_\mathrm{H}(s')\;\mathrm{d}s',
\end{equation}
where
\begin{equation}
    N_\mathrm{c}(s) = \frac{\mu_0 {E_\mathrm{c}}^2}{6\pi e^4 \gamma(s)}.
\end{equation}
The quantity $\tau^*(s)$ is defined analogously to $\tau(s)$, but for a fully ionised plasma where $\gamma(s)$ equals the Coulomb logarithm $\ln\Lambda$:
\begin{equation}
    \label{eq:tau_star}
    \tau^*(s) = \frac{N^*(s)}{N^*_\mathrm{c}} = \frac{1}{N^*_\mathrm{c}}\int_0^s \frac{\gamma(s')}{\ln\Lambda}n_\mathrm{H}(s')\;\mathrm{d}s',
\end{equation}
where
\begin{equation}
    \label{eq:stopping_depth_star}
    N^*_\mathrm{c} = \frac{\gamma(s)}{\ln\Lambda}N_\mathrm{c}(s).
\end{equation}

To determine the power $\Delta\mathcal{E}$ deposited by the beam over a field line segment from $s$ to $s + \Delta s_\mathrm{dep}$, we need to integrate Eq. \eqref{eq:power_per_dist} over the segment:
\begin{equation}
    \label{eq:power_integral}
    \Delta\mathcal{E} = \int_s^{s + \Delta s_\mathrm{dep}} \frac{\mathrm{d}\mathcal{E}}{\mathrm{d}s}(s')\;\mathrm{d}s' = \int_0^{\Delta s_\mathrm{dep}} \frac{\mathrm{d}\mathcal{E}}{\mathrm{d}s}(s + \delta s)\;\mathrm{d}\delta s.
\end{equation}
This integral must be evaluated with sufficient accuracy never to significantly violate energy conservation. We can achieve this in a computationally efficient manner by observing that there will be little variation in the local plasma properties along the segment when we set $\Delta s_\mathrm{dep}$ much smaller than a grid cell. In this case, we can assume that $n_\mathrm{H}(s + \delta s) \approx n_\mathrm{H}(s)$ and $\gamma(s + \delta s) \approx \gamma(s)$, which gives (using Eq. \eqref{eq:tau})
\begin{equation}
    \tau(s + \delta s) \approx \tau(s) + \frac{n_\mathrm{H}(s)}{N_\mathrm{c}(s)}\delta s
\end{equation}
and (using Eqs. \eqref{eq:tau_star} and \eqref{eq:stopping_depth_star})
\begin{equation}
    \tau^*(s + \delta s) \approx \tau^*(s) + \frac{\gamma(s)}{\ln\Lambda}\frac{n_\mathrm{H}(s)}{N^*_\mathrm{c}}\delta s = \tau^*(s) + \frac{n_\mathrm{H}(s)}{N_\mathrm{c}(s)}\delta s.
\end{equation}
Defining
\begin{equation}
    \delta\tau(s) = \frac{n_\mathrm{H}(s)}{N_\mathrm{c}(s)}\delta s \quad\mathrm{and}\quad \Delta\tau(s) = \frac{n_\mathrm{H}(s)}{N_\mathrm{c}(s)}\Delta s_\mathrm{dep},
\end{equation}
we can approximate the integral in Eq. \eqref{eq:power_integral} as
\begin{equation}
    \label{eq:simplified_power_integral}
    \Delta\mathcal{E} \approx \int_0^{\Delta\tau} C N_\mathrm{c}\gamma B\left(\mathrm{min}\left(\tau + \delta\tau, 1\right); \frac{\delta}{2}, \frac{1}{3}\right)\left(\tau^* + \delta\tau\right)^{-\delta/2}\;\mathrm{d}\delta\tau.
\end{equation}
For $\tau(s) \geq 1$, the incomplete beta function $B$ becomes independent of $\delta\tau$, enabling analytical solution of the integral:
\begin{equation}
    \Delta\mathcal{E}(\tau \geq 1) \approx C N_\mathrm{c}\gamma B\left(1; \frac{\delta}{2}, \frac{1}{3}\right)\frac{{\tau^*}^{1 - \delta/2} - \left(\tau^* + \Delta\tau\right)^{1 - \delta/2}}{\delta/2 - 1}.
\end{equation}
For $\tau(s) + \delta\tau < 1$, Eq. \eqref{eq:simplified_power_integral} must be evaluated numerically. Since the integrand is relatively smooth, it is well approximated by a low-order polynomial. In our implementation, we therefore perform the integration using a 3-point Gauss--Legendre quadrature, which provides sufficient accuracy while remaining computationally cheap.

\subsection{Parallelisation}

\begin{figure*}[!thb]
    \includegraphics{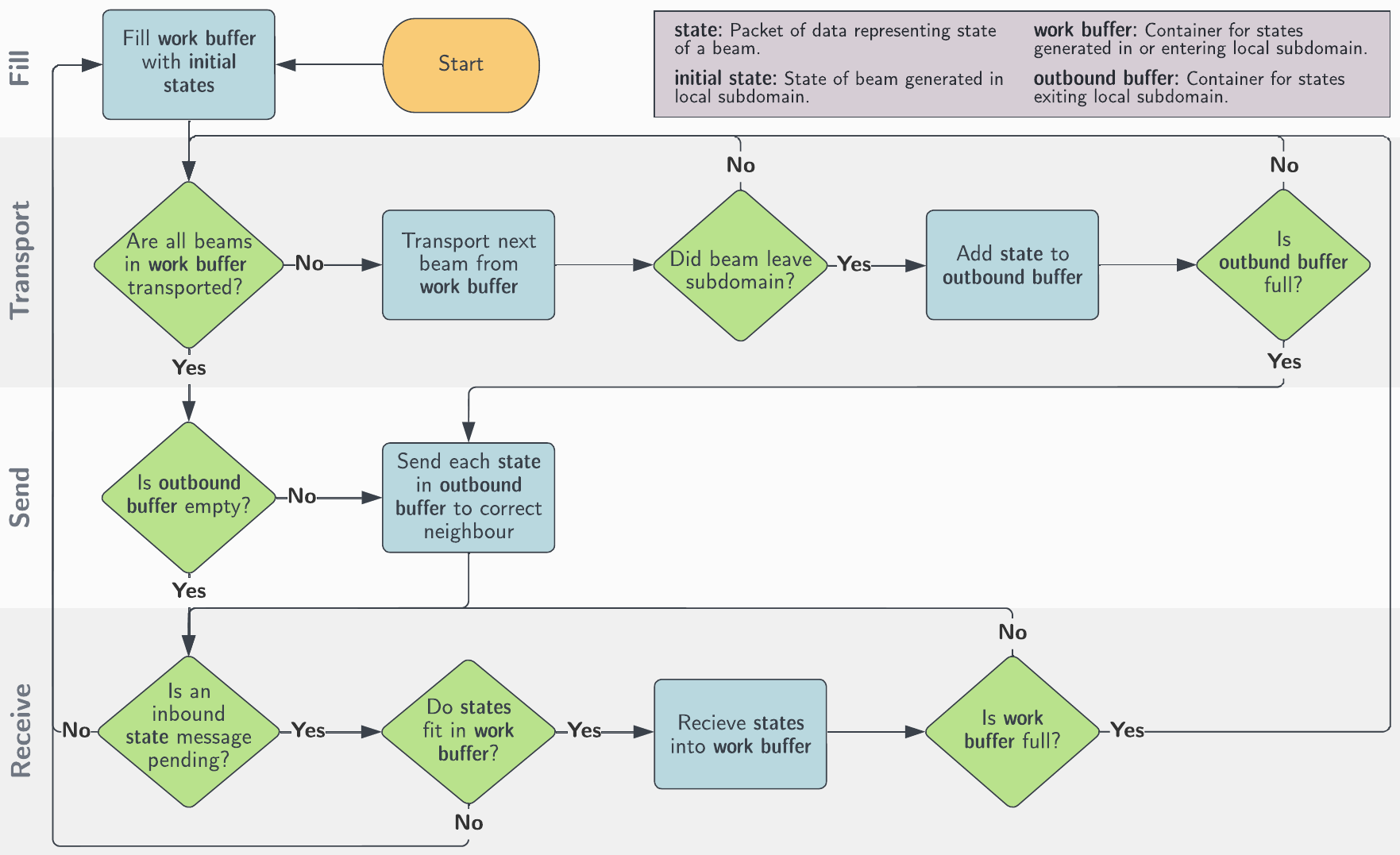}
    \centering
    \caption{Flowchart outlining the logic each process uses for managing beam transport and communication of beams with neighbouring processes.}
    \label{fig:tracing_communication}
\end{figure*}

The Bifrost code is designed to run in parallel on a distributed memory system with many CPU cores \citep{hayek_radiative_2010, gudiksen_stellar_2011}, with each core executing the code as a separate process. Bifrost divides the full simulation domain into a grid of subdomains of equal size, one subdomain for each process. Each process runs the physical simulation only for its subdomain, as the field values within the subdomain are the only ones available in its private memory. At every time step, the process synchronises its boundary values with the processes responsible for the neighbouring subdomains. This synchronisation uses the Message Passing Interface (MPI) \citep{message_passing_interface_forum_mpi_2008} for communicating values between processes.

The domain decomposition parallelism employed in Bifrost has major implications for the simulation of energy transport by electron beams. Because the instantaneous propagation of a beam may follow any trajectory through the whole simulation domain, we must be able to trace each beam trajectory through an arbitrary sequence of subdomains at each time step. Doing this in a manner that minimises both idle time and communication overhead requires that we carefully coordinate the passing of beam state between processes.

The three main tasks a process has to manage are tracing beams through its subdomain, sending the state of each outbound beam to the appropriate process, and receiving the states of inbound beams. Figure \ref{fig:tracing_communication} shows how we coordinate these tasks in our implementation. Each process starts by tracing some of the beams they generated in their subdomain. It stores the states of traced beams leaving the subdomain in a buffer, grouped according to the neighbour for which they are destined. This buffer is referred to as the `outbound buffer' in Fig. \ref{fig:tracing_communication}. When the process sends the states in the outbound buffer, it dispatches a single message containing a group of beams to each neighbour. Sending states in groups reduces the communication overhead compared to sending individual states. However, because it will make the process send states less frequently than if it were to send them individually, it does increase the risk of neighbouring processes being idle because they are waiting for work. To ensure a steady throughput of beams, each process always sends its outbound states whenever it has traced a certain number of beams, regardless of how populated the outbound buffer may be. After sending its outbound beams, the process obtains more work by receiving inbound groups of beams from its neighbours. It places the received states into a buffer whose size corresponds to the maximum number of beams to trace between sending the outbound buffer. This buffer is referred to as the `work buffer' in Fig. \ref{fig:tracing_communication}. If the process has received all pending messages without filling the work buffer, it tries to populate the remaining space with locally generated beam states. In this way, the process prioritises inbound beams while keeping the amount of work between communications as steady as possible. It then traces all the beams in the work buffer before again sending any outbound beams and repeating the cycle.

In addition to tracing and communicating beams, the processes need to determine when all beams are completed to know when to exit the cycle. For clarity, we omit this logic in Fig. \ref{fig:tracing_communication}, but our approach is straightforward. A designated process obtains the total number of generated beams through collective communication between the processes as soon as they have generated every beam. As the processes trace beams, they keep a count of the number they have completed within their subdomain. Whenever a process becomes idle, it reports the count to the designated process, which sends out a collective termination message once all beams are completed.

\section{Results}
\label{sec:results}

\subsection{Tracing accuracy}
\label{sec:tracing_analysis}

\begin{figure}[!thb]
    \includegraphics{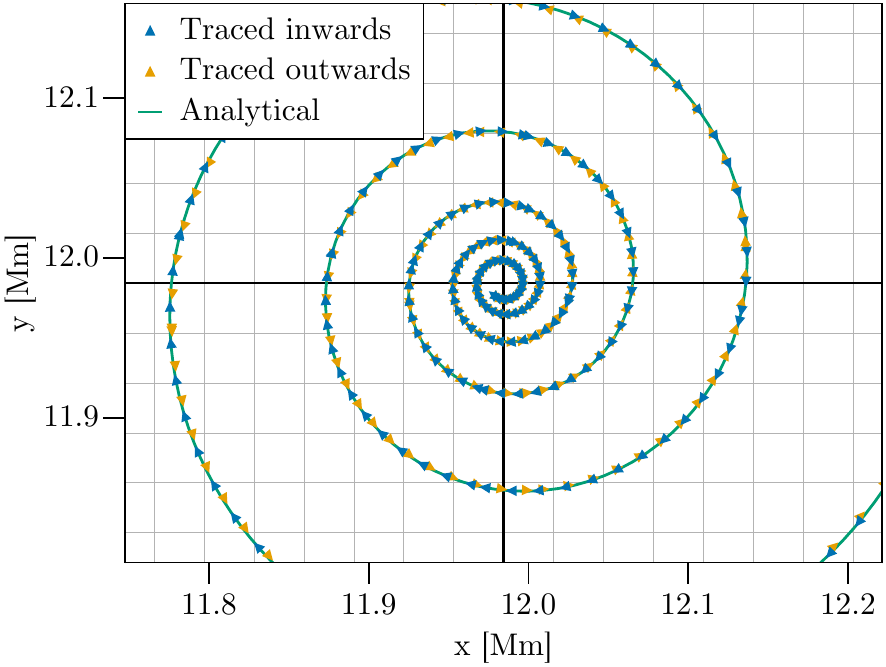}
    \centering
    \caption{Field line traced in the vector field given by Eq. \eqref{eq:spiral_vector_field} with $k = 0.1$ together with the corresponding analytical field line. We began tracing at a position offset by 1 Mm from the spiral centre in the $x$-direction and followed the spiral trajectory inwards until the distance to the centre reached 0.01 Mm before reversing and tracing the same distance in the opposite direction. Each arrowhead shows the location of a tracing step and points in the direction that the step was taken. Blue arrowheads depict the inward trajectory, and orange arrowheads depict the outward trajectory. The green curve is the corresponding analytically determined field line, a logarithmic spiral. The grey grid in the background represents the grid of the Bifrost simulation, while the thicker black lines are the boundaries between subdomains owned by separate processes.}
    \label{fig:spiral}
\end{figure}

To judge the accuracy of our parallelised implementation of field line tracing, we performed tests where we compared field lines traced in a simple vector field to analytical solutions. We designed the tests to emulate a challenging scenario where a beam starts in a region with a smooth magnetic field and passes through a region with strongly varying magnetic field directions before returning to the smooth region. For the tests, we replaced the magnetic field in the simulation with the vector field
\begin{align}
    \label{eq:spiral_vector_field}
    \mathbf{V}(x, y, z) & = \begin{pmatrix}
                                k(x - x_\mathrm{c}) - (y - y_\mathrm{c}) \\
                                k(y - y_\mathrm{c}) + (x - x_\mathrm{c}) \\
                                0
                            \end{pmatrix},
\end{align}
whose field lines are logarithmic spirals in the $xy$-plane centred on $(x, y) = (x_\mathrm{c}, y_\mathrm{c})$, with polar slopes $k$. We then traced field lines in the inward direction, circling the spiral centre at gradually smaller distances, before reversing direction and pursuing the same trajectory out again once the field line came very close to the centre. An example of a traced line together with its analytical counterpart is shown in Fig. \ref{fig:spiral}. To evaluate the performance of the adaptive step length control method outlined in Sect. \ref{sec:tracing}, we performed such tests using both adaptive stepping and a fixed step length.

\begin{figure*}[!thb]
    \includegraphics{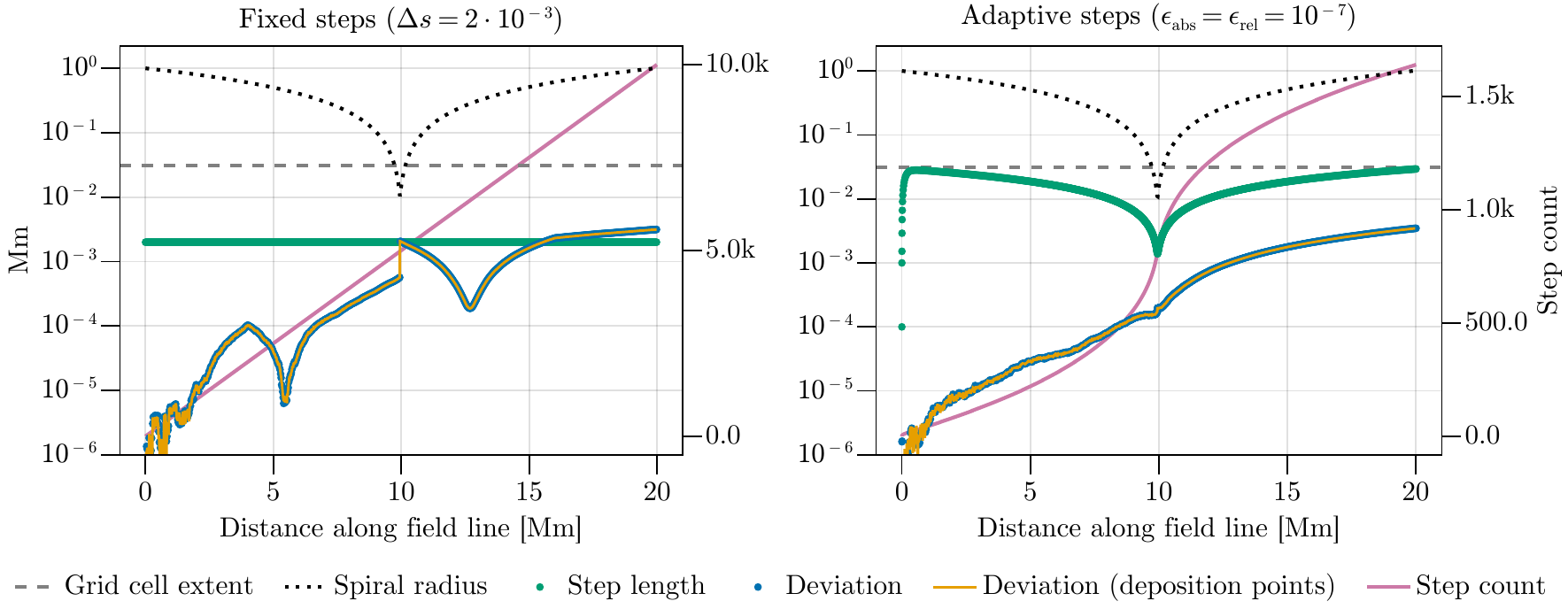}
    \centering
    \caption{Deviations of a traced field line from the corresponding analytical field line as a function of traced distance, using a fixed (left-hand side) and an adaptive (right-hand side) stepping scheme. The vector field is the same as the one used for Fig. \ref{fig:spiral}. The black dotted curves show how the distance to the centre of the spiral changes with traced distance, while the grey dashed lines indicate the horizontal extent of a grid cell for reference. The blue dots show the deviations $\left\lVert\mathbf{x}(s) - \tilde{\mathbf{x}}(s)\right\rVert$ of the positions $\mathbf{x}(s)$ produced by the stepping scheme from the corresponding positions $\tilde{\mathbf{x}}(s)$ of the analytical field line. Similarly, the orange curves overlapping the blue dots show the deviations of the regularly spaced deposition points found by interpolating between the stepping positions. Green dots indicate the length of each step along the field line. We note that the abrupt increase in step length at the start of the adaptively traced field line stems from our conservative choice of $10^{-4}\;\mathrm{Mm}$ as the initial step length. Finally, the red curves show how the number of steps (measured on the right-hand side axes) accumulates as the field lines are traced.}
    \label{fig:spiral_tracing_verification}
\end{figure*}

Our tests show that we can accurately trace the spiral field line using both a fixed and adaptive step length. This is evident from Fig. \ref{fig:spiral_tracing_verification}, which contains plots of the deviation of the traced field line from the analytical field line as a function of traced distance for two tests with fixed and adaptive steps. The deviations are the same when we measure them for the regularly spaced deposition points rather than the points being stepped to, showing that we determine the deposition points with acceptable accuracy. We chose the tolerance for the adaptive stepping in Fig. \ref{fig:spiral_tracing_verification} to demonstrate that we can trace the spiral tens of megametres through varying degrees of curvature without deviating more than a fraction of a grid cell extent from the analytical solution. We then selected a fixed step length that yields a field line with accuracy similar to the adaptive stepping.

Even though fixed steps can yield the same accuracy as adaptive steps, this requires significantly more computation. The adaptive stepping scheme uses much longer steps than the fixed stepping scheme in the outer parts of the spiral, only shortening them below the fixed step length to maintain accuracy as the spiral radius shrinks below grid cell scales. This is clear from the points indicating step lengths in Fig. \ref{fig:spiral_tracing_verification}. The result is that adaptive stepping reduces the required number of steps by nearly an order of magnitude compared to fixed stepping in this specific situation, as seen from the step count curves in Fig. \ref{fig:spiral_tracing_verification}.

Our tests also demonstrate that multiple processes can correctly communicate the state of traced beams between them as the beams pass between subdomains. This is because they trace the field line with continuously high accuracy even though, as shown in Fig. \ref{fig:spiral}, we placed the spiral centre at a point where four subdomains belonging to separate processes meet, requiring repeated crossing between subdomains.

\subsection{Parallel scalability}
\label{sec:scalability}

\begin{figure}[!thb]
    \includegraphics{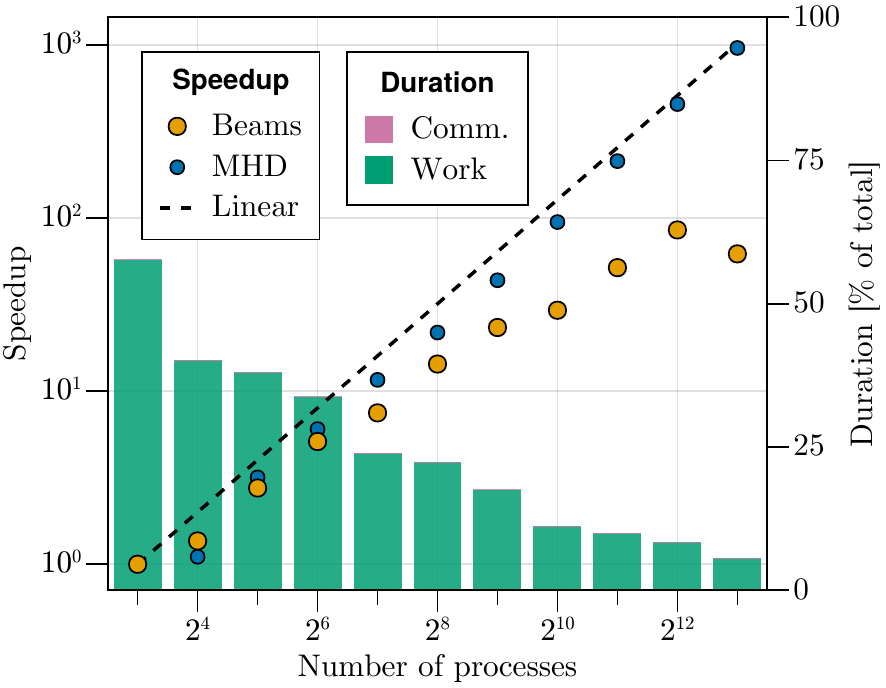}
    \centering
    \caption{Scalability of the parallel execution of electron beam transport. We performed all runs on Sigma2's Betzy supercomputer. The orange dots show the measured speedup (the relative increase in execution speed with respect to the slowest execution) of electron beam transport for process counts ranging from 8 to 8192. For reference, the corresponding speedups of Bifrost's MHD solver are indicated as blue dots, along with a dashed line representing linear speedup. Each green bar shows the average percentage of the execution duration of the electron beam transport that the processes spend performing useful work (generating and transporting beams) for a given process count. Stacked on top are red bars indicating the corresponding percentage the processes spend communicating with each other, but these are too small to be visible.}
    \label{fig:speedup}
\end{figure}

We measured the scalability of our parallel implementation of electron beam transport by timing runs executed with a range of different process counts and calculating the speedups relative to the slowest execution time. For comparison, we also measured the speedups of the part of Bifrost that solves the MHD equations. Figure \ref{fig:speedup} shows the results. The MHD solver exhibits close to linear speedup, meaning that doubling the number of processes halves the execution time. This indicates that the MHD solver can distribute its work evenly between numerous processes with minimal communication overhead. The electron beam transport, on the other hand, does not attain linear speedup. Instead, the relative decrease in execution time is smaller than the relative increase in process count, and this deviation becomes larger as we add processes.

The reason for the non-optimal speedup is that each process, on average, spends a smaller proportion of its execution time doing useful work. As the bars in Fig. \ref{fig:speedup} show, the processes in an 8-process run spend about $60\%$ of the execution time generating and transporting beams on average. In a run with 8192 processes, this number is only $6\%$. Since it is clear from the figure that communication overhead is negligible, the processes spend the remainder of the execution time in an idle state with no work to do.

\begin{figure}[!thb]
    \includegraphics{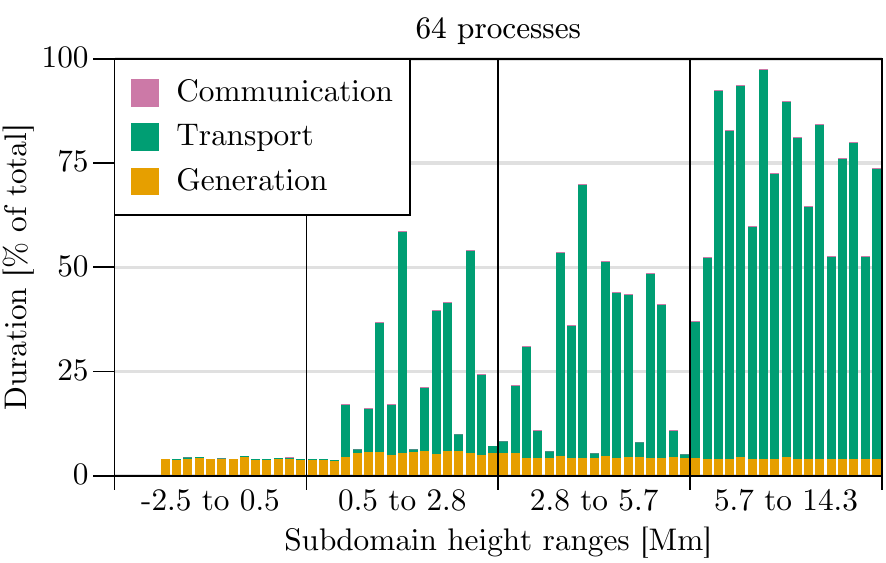}
    \centering
    \caption{Proportions of the total time of an electron beam transport execution spent performing specific tasks by each process in a run with 64 processes. Each bar represents a single process and consists of three stacked parts. The bottom part (orange) indicates the percentage of the time spent generating electron beams, the middle part (green) the time spent transporting them and the top part (red) the time spent communicating beams with neighbours. The red parts representing communication are generally too small to be visible. On the horizontal axis, the processes are grouped according to the heights that their subdomains span in the atmosphere. With 64 processes, there are $4 \times 4 \times 4$ subdomains, hence 4 height groups, each with 16 processes. We note that the height ranges differ in extent because the simulation grid has irregular grid cell heights.}
    \label{fig:timings}
\end{figure}

To better understand the cause of the high degree of idleness, we measured the time spent on beam generation, transport, and communication by the individual processes. As apparent in Fig. \ref{fig:timings}, the measurements reveal that the processes -- while all unencumbered by communication overhead -- experience a substantial imbalance in workload. The processes whose subdomains comprise the top of the simulated atmosphere generally spend most of their time transporting beams. Lower in the atmosphere, most processes have slightly more work generating beams than higher up (because most beams originate close to the transition region) but significantly less work transporting them. These processes thus spend most of their time simply waiting for the busiest process to complete its work.

A potential source of idleness is a too slow flow of information through the system. Too infrequent communication of beams across subdomain boundaries would leave processes with potential work to do waiting for it longer than necessary. In our test runs, this appears not to be the case. The time the busiest process takes to complete its work is only slightly shorter than the total execution time for electron beam transport. Thus, it does not have to wait long after sending its last beams for the other processes to complete them.

\section{Discussion and conclusions}
\label{sec:discussion}

The numerical implementation of electron beam transport presented here enables us to robustly compute the trajectories of beams and integrate the energy deposited along them. The tracing tests explained in Sect. \ref{sec:tracing_analysis} confirm that we can trace a beam trajectory through a multi-scale magnetic field in the domain decomposed simulation box with insignificant deviation from the analytical field line. Moreover, we can do so in a minimal number of steps by using an embedded Runge--Kutta scheme combined with error-based step length adjustment. We also interpolate between the steps to obtain deposition points at regular intervals that are small compared to the extent of grid cells. As a result, we can evaluate the rate of energy deposition into each grid cell efficiently and accurately by integrating Eq. \eqref{eq:power_per_dist} between points under the assumption of uniform plasma conditions.

The timing measurements presented in Sect. \ref{sec:scalability} show that the way we coordinate transport and communication of beams leads to an efficient flow of information between processes. By buffering traced beams and sending them to their destined processes in groups, we reduce the number of required messages and achieve negligible communication overhead. Furthermore, by making sure that we do not buffer outbound beams for too long before sending them, we prevent unnecessary waiting for work.

\begin{figure}[!thb]
    \includegraphics{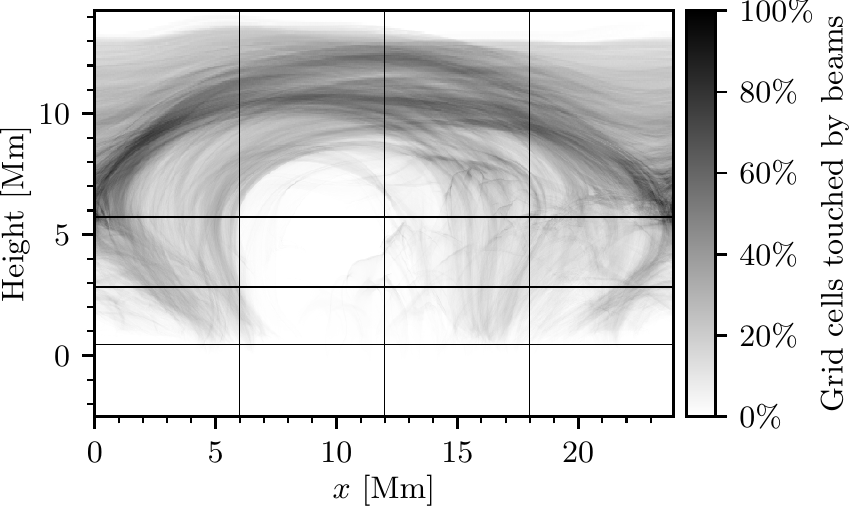}
    \centering
    \caption{Amount of electron beam activity in the simulation used for measuring parallel scalability. The shade of black at each point indicates the percentage of the grid cells along the $y$-axis of the simulation snapshot that a beam has passed through. The straight black lines are the subdomain boundaries in a run with 64 processes.}
    \label{fig:beam_presence}
\end{figure}

Despite efficient communication, our implementation exhibits less than ideal scalability. Increasing the number of processes from 8 to 8000 only provides a speedup of about 100. The reason is a considerable imbalance of total workload between the processes. The imbalance happens because electron beam transport only takes place in a fraction of the total volume of the simulation domain, as evident from Fig. \ref{fig:beam_presence}. For the particular simulation setup used in this work (the same as presented in \citetalias{frogner_accelerated_2020}), only about $10\%$ of the grid cells are affected by electron beam transport. As more processes are used, the simulation domain gets decomposed into smaller subdomains, and a larger proportion of the subdomains experience little or no electron beam activity.

Bifrost's use of varying grid cell heights that are smallest around the transition region and larger in the corona increases the workload imbalance further. It causes subdomains in the corona to cover larger physical volumes than subdomains lower in the atmosphere, even though they have the same number of grid cells. This is evident from the subdomain boundaries shown in Fig. \ref{fig:beam_presence}. Since most of the electron beam transport occurs in the corona, this produces more work for the processes that already handle the busiest regions.

An imbalance of workload between processes is an inherent issue in using domain decomposition to parallelise a computation with an uneven spatial distribution of work. Other parallel programming models, such as shared memory or a partitioned global address space, may arguably be better suited for the isolated problem of transporting large numbers of electron beams. However, when transport is performed as part of a full atmospheric simulation, as required for the electron beams to influence the evolution of the atmosphere, the implementation has to conform to the parallelisation method the rest of the simulation employs.

One possible approach for balancing the load of transporting electron beams is allowing a process to fork additional threads, depending on its demand, for transporting its beams concurrently, using an interface for shared memory multiprocessing, such as OpenMP. The actual number of  CPU cores executing these support threads would depend on the number of available cores on the compute node running the process. For simulations where electron beam transport is the slowest component, it could be beneficial for the execution time as a whole to reduce the total number of processes and free up cores for executing support threads on demand.

\begin{acknowledgements}
    This research was supported by the Research Council of Norway through its Centres of Excellence scheme, project number 262622, and through grants of computing time from the Programme for Supercomputing.
\end{acknowledgements}

\bibliographystyle{aa}
\bibliography{references.bib}

\end{document}